\documentclass[12pt]{iopart}
\usepackage{graphicx,epstopdf}
\let\text\textrm

\begin{document}
\title{Laval nozzle as an acoustic analogue of a massive field}
\author{M. A. Cuyubamba}
\address{Universidade Federal do ABC (UFABC), Rua Aboli\c{c}\~ao, CEP: 09210-180, Santo Andr\'e, SP, Brazil}
\ead{marco.espinoza@ufabc.edu.br}
\begin{abstract}

We study a gas flow in the Laval nozzle, which is a convergent-divergent tube that has a sonic point in its throat. We show how to obtain the appropriate form of the tube, so that the acoustic perturbations of the gas flow in it satisfy any given wavelike equation. With the help of the proposed method we find the Laval nozzle, which is an acoustic analogue of the massive scalar field in the background of the Schwarzschild black hole. This gives us a possibility to observe in a laboratory the quasinormal ringing of the massive scalar field, which, for special set of the parameters, can have infinitely long-living oscillations in its spectrum.

\end{abstract}

\section{Introduction}

Massive fields in the vicinity of black holes have been studied during the last two decades (see \cite{Konoplya:2011qq} for review). It was found that their behaviour is qualitatively different from the behaviour of the massless fields. The response of a black hole upon any perturbations at late times can be described by a characteristic spectrum of exponentially damped oscillations. The spectrum of a massive field, for some particular values of the parameters, has the oscillations with a very small decay rate in their characteristic spectra. These oscillations, that behave similarly to standing waves, were called quasiresonances \cite{Konoplya:2004wg,Ohashi:2004wr}. The asymptotical behaviour of massive fields is also different: one observes the oscillating tails that decay as inverse power of time, which is universal at the asymptotically late times \cite{Konoplya:2006gq}.

Yet, since massive fields are short-ranged, we cannot expect the observation of their signal from black holes in near-future experiments. An attractive possibility for experimental study of the massive fields in the background of a black hole is a consideration of the acoustic analogue. This is a well-known Unruh analogue of a black hole \cite{Unruh}, which is an inhomogeneous fluid system, where the perturbations (sound waves) can be described by a Klein-Gordon equation in the background of some effective curved metric \cite{3}.

The sound waves in a fluid can propagate from a subsonic region to a supersonic one, but they cannot go back. Therefore, sonic points in a fluid with a space-dependent velocity form a one-way surface for the sound waves, which is called ``acoustic horizon'' by similarity with the event horizon of the black hole.

The works of Unruh stimulated the study of various acoustic systems, such as:
\begin{enumerate}
\item the ``draining bathtub'' which is an analogue of a rotating black hole  \cite{5,6,6.5};
\item the Bose-Einstein condensate \cite{7,8,9,10}, which, in the regime when the thermal fluctuations can be neglected, allows to observe a phonon analogue of the Hawking radiation \cite{11,12,13,14,15,16};
\item the so-called optical black holes \cite{17} due to sound waves in a photon fluid of an optical cavity and inside an optical fiber \cite{18}; and others \cite{19,20,21}.
\end{enumerate}
Within the analogue-gravity approach one considers hydrodynamical equations as field equations in some effective background which is not a solution to Einstein equations. Using this approach the one-dimensional perturbations in the Laval nozzle were studied in \cite{22}. It was found that the perturbations of the gas flow in the Laval nozzle can be described by a wave-like equation with the effective potential, which depends on the form of the tube. The inverse problem for the correspondence of the form of the Laval nozzle to the Schwarzschild black holes has been solved in \cite{23}, where the form of the Laval was found in order to obtain acoustic analogue for the perturbations of massless fields.

Here we describe a method, which allows us to find an appropriate form of the Laval nozzle for any given effective potential. We use this method to obtain the acoustic analogue of the massive scalar field in the background of the Schwarzschild black hole. This paper is organized in the following form: In the section II we give the basic equations for a one-dimensional flow in the Laval nozzle and its perturbations. In the section III we describe the numerical method, which allows us to find the appropriate nozzle form in order to mimic any given effective potential. In the section IV we apply the method to find the form of the Laval nozzle, which is an acoustic analogue of the massive scalar field in the Schwarzschild background and show the corresponding time-domain profiles. Finally, in the conclusion, we discuss the obtained results and open questions.

\section{Basic equations}

A perfect fluid in the Laval nozzle can be described by the continuity equation and the Euler equation, that read, respectively,
\begin{eqnarray}
\partial_{t}(\rho A)+\partial_x (\rho \upsilon A)=0,\label{forma1}\\
\rho\left(\partial_t+\vec\upsilon\cdot\nabla\right)\vec\upsilon=-\nabla p,\label{forma2}
\end{eqnarray}
where $\rho$ is the density of a gas, $\vec\upsilon$ is the fluid velocity, $p$ is the pressure, and $A$ is the cross-section area of the nozzle. Following \cite{22} we assume that the fluid is isentropic and the pressure depends only on the density
\begin{equation}
p\propto\rho^\gamma,\label{forma3}
\end{equation}
where $\gamma$ is the heat capacity ($\gamma=1.4$ for the air).

Assuming that the flux is irrotational $\nabla\times\vec\upsilon=\vec 0$, the velocity can be expressed as $\vec\upsilon=\nabla\Phi$, where $\Phi=\int\upsilon dx$ is the velocity potential, which satisfies the Bernoulli equation

\begin{equation}
\partial_t\Phi+\frac{1}{2}\left(\partial_x\Phi\right)^2+h(\rho)=0.\label{forma5}
\end{equation}

We study linear perturbations of the flux, i.e. we consider the fluid density $\rho$ and the velocity potential $\Phi$ as
\begin{eqnarray}
\rho&=\bar{\rho}+\delta\rho\text{,}\quad\bar{\rho}\gg|\delta\rho|\label{forma6}\\
\Phi&=\bar{\Phi}+\phi\text{,}\quad|\partial_x \bar{\Phi}|\gg|\partial_x \phi|\label{forma7}
\end{eqnarray}
where $\bar{\rho}$, $\bar{\Phi}$ are the background dynamical quantities which satisfy (\ref{forma1}) and (\ref{forma5}), $\delta\rho$ and $\phi$ describe the perturbations, which are considered small so that we neglect the higher-order corrections.

We introduce the function $H_\omega (x)$,
\begin{equation}
H_\omega (x)=g^{1/2} \int dt e^{i\omega(t-a(x))} \phi (t,x),\label{forma8}
\end{equation}
with
\begin{equation}
g=\frac{\rho A}{c_s},\qquad
a(x)=\int\frac{|\upsilon| dx}{c_s^2-\upsilon^2},\label{forma11}
\end{equation}
where $c_s$ is the sound speed,
\begin{equation}
c_s=\sqrt{\frac{dp}{d\rho}}=\sqrt{\frac{\gamma p}{\rho}}.
\end{equation}

We find that $H_\omega$ satisfies the Schr\"{o}dinger-type wave-like equation
\begin{eqnarray}
&\left(\frac{d^2}{dx^{*2}}+\kappa^2-V(x^*)\right)H_\omega(x^*)=0\label{forma12}\\
&V(x^*)=\frac{1}{g^2}\left[\frac{g}{2}\frac{d^2g}{dx^{*2}}-\frac{1}{4}\left(\frac{dg}{dx^*}\right)^2\right]\label{forma13}
\end{eqnarray}
with respect to the new variable
\begin{equation}
x^*=\int\frac{c_{s0}c_{s}dx}{c_s^2-\upsilon^2},\label{forma9}
\end{equation}
where $\kappa=\omega/c_{s0}$ and $c_{s0}$ is the stagnation sound speed.

The coordinate $x^*$ is the tortoise coordinate for the analogue black hole: $x^*=-\infty$ at the throat and $x^*=\infty$ corresponds to the spatial infinity ($x=\infty$).

Following \cite{23}, we measure $A$ and $\rho$, respectively, in the units of cross-sectional area at the throat ($A_{th}$) and the flux stagnation density ($\rho_0$), and choose the arbitrary factor for the function $g$ in such a way that
\begin{equation}
g=\frac{\rho A}{2\rho^{(\gamma-1)/2}},\quad A^{-1}=\left(1-\rho^{(\gamma-1)}\right)^{1/2}\rho.\label{forma15}
\end{equation}
Then the cross section area can be expressed as a function of $g$ as
\begin{equation}
A=\frac{\sqrt{2}\left(2g^2 \left(1-\sqrt{1-g^{-2}}\right)\right)^{1/(\gamma-1)}}{\sqrt{1-\sqrt{1-g^{-2}}}}.\label{forma16}
\end{equation}
We find also that
\begin{equation}
\frac{\upsilon^2}{c_s^2}=\frac{2}{\gamma-1}\left(2g^2\left(1-\sqrt{1-g^{-2}}\right)-1\right).\label{forma16.5}
\end{equation}
Since the gas velocity is equal to the sound velocity at the acoustic horizon, we obtain
\begin{equation}
g \Big|_{horizon}=\frac{\gamma+1}{2\sqrt{2}\sqrt{\gamma-1}}=\frac{3}{\sqrt{5}}.\label{forma17}
\end{equation}

\section{The nozzle form from a given effective potential}

Linear perturbations of a spherically-symmetric black hole, after decoupling of the time and angular variables, can always be reduced to the following wave-like equation
\begin{equation}
\left(\frac{d}{dr^{*2}}+\omega^2-V(r^*)\right)R(r^*)=0,\label{forma18}
\end{equation}
where the effective potential $V=V(r^*)$ depends on the parameters of the field and the black hole and the tortoise coordinate is defined as
\begin{equation}
r^*=\int\frac{dr}{f(r)},
\end{equation}
where $f(r)$ depends on the parameters of the black hole.

In order to find the form of the Laval nozzle which is an acoustic analogue of the black hole perturbations we equate the tortoise coordinates and the effective potentials of the equations (\ref{forma12}) and (\ref{forma18})
\begin{equation}
\frac{f(r)f'(r)g'(r)+f(r)^2g'(r)^2}{2g(r)}-\frac{f(r)^2g'(r)^2}{4g(r)^2}=V(r).\label{forma19}
\end{equation}
From $dx^*=dr^*$ and equations (\ref{forma16.5}) and (\ref{forma9}) we find the relation between the coordinate of the nozzle and the radial coordinate of the metric $r$:
\begin{equation}
dx=\frac{\left(\gamma+1-4g(r)^2\left(1-\sqrt{1-g(r)^{-2}}\right)\right)dr}{f(r)(\gamma-1)\sqrt{2g(r)^2\left(1-\sqrt{1-g(r)^{-2}}\right)}}.\label{forma26}
\end{equation}
If $g(r)$ is known, from the equations (\ref{forma16}) and (\ref{forma26}), one can find the function $A(x)$, which describes the nozzle form.

In order to find $g(r)$ we make the substitution $g(r)=h(r)^2$. Then the differential equation (\ref{forma19}) reads
\begin{equation}
f(r)^2h''(r)+f(r)f'(r)h'(r)-V(r)h(r)=0.\label{forma21}
\end{equation}

Since the function $f(r)$ vanishes at the event horizon $r=r_+$, the linear equation (\ref{forma21}) always has a regular singular point there. Using the Frobenius method we expand the general solution to the differential equation near the event horizon as
\begin{equation}
h(r)=c_1h_1(r) + c_2h_2(r),\label{forma22}
\end{equation}
where $c_1$ and $c_2$ are arbitrary constants,
\begin{equation}
h_1(r)=(r-r_+)^{\lambda_1}\left(1+\sum_{n=1}^\infty a_n(r-r_+)^n\right),\label{forma23}
\end{equation}
$$h_2(r)=h_1(r)\ln(r-r_+)+(r-r_+)^{\lambda_2}\sum_{n=0}^\infty b_n(r-r_+)^n,$$
when $\lambda_1-\lambda_2$ is an integer, and
$$h_2(r)=(r-r_+)^{\lambda_2}\left(1+\sum_{n=1}^\infty b_n(r-r_+)^n\right),$$
otherwise, $\lambda_2\leq\lambda_1$ are the roots of the indicial equation and depend on the given functions $f(r)$ and $V(r)$.

In order to satisfy (\ref{forma17}), one of the roots must be zero. $f'(r_+)>0$ implies that the other root is negative. Hence, for $\lambda_2\leq\lambda_1=0$, $h_2(r)$ is always divergent at the horizon $r=r_+$ and we choose $c_2=0$. Therefore, from (\ref{forma17}) we find that
$$c_1=\sqrt{\frac{\gamma+1}{2\sqrt{2}\sqrt{\gamma-1}}}=\sqrt{\frac{3}{\sqrt{5}}}.$$
We expand (\ref{forma23}) near the event horizon and find $h'(r+)$, which completely fixes the initial value problem at $r=r_+$. Then, we are able to solve numerically the equation (\ref{forma21}) using the Runge-Kutta method for $r>r_+$.

\section{Acoustic analogue for the massive scalar field}
We consider the massive scalar field in the background of the Schwarzschild black hole, given by the line element
\begin{equation}\label{metric}
ds^2 = f(r)dt^2-\frac{dr^2}{f(r)}-r^2(d\theta^2+\sin^2\theta d\phi^2),\quad f(r)=1-\frac{2M}{r},
\end{equation}
where $M$ is the mass of the black hole. Hereafter we measure all the quantities in units of the black hole horizon, i.e. $r_+=2M=1$.

The scalar field $\Psi$ satisfies the Klein-Gordon equation
\begin{equation}
\left(\nabla^\mu\nabla_\mu + m^2\right)\Psi=0,\label{forma27}
\end{equation}
where $\nabla_\mu$ is the covariant derivative, $m$ is the field mass. The equation (\ref{forma27}) in the background (\ref{metric}) reads
\begin{equation}
\frac{1}{\sqrt{|g|}}\partial_\mu\left(g^{\mu\nu}\sqrt{|g|}\partial_\nu\Psi\right)+m^2\Psi=0.\label{forma28}
\end{equation}
After the separation of the angular and time variables
\begin{equation}
\Psi(t,r,\theta,\phi)=\frac{R(r)}{r}Y^m_l (\theta,\phi)e^{-i\omega t}\label{forma29}
\end{equation}
we obtain the wave-like equation (\ref{forma18}) with the effective potential
\begin{equation}
V(r)=f(r)\left(\frac{l(l+1)}{r^2}+\frac{f'(r)}{r}+m^2\right).\label{forma30}
\end{equation}

In order to show the time-domain evolution of perturbations we use the discretization scheme proposed by Gundlach, Price, and Pullin \cite{31}. We consider the time-dependent equation
\begin{equation}
\left(\frac{d^2}{d{r^*}^{2}}-\frac{d^2}{dt^2}-V(r)\right)\Phi(t,r^*)=0.\label{forma32}
\end{equation}

Rewriting (\ref{forma32}) in terms of the light-cone coordinates $du=dt-dr^*$ and $dv=dt+dr^*$, we find that
\begin{equation}
\Phi(N)=\Phi(W)+\Phi(E)-\Phi(S)-\frac{h^2}{8}V(S)\left[\Phi(W)+\Phi(E)\right]+\mathcal{O}(h^4),\label{forma35}
\end{equation}
where the point $N$, $M$, $E$ and $S$ are the points of one square in a grid with step $h$ in the $u$-$v$ plane, as follows: $S=(u,v)$, $W=(u+h,v)$, $E=(u,v+h)$ and $N=(u+h,v+h)$. With the initial data specified on two null-surfaces $u = u_0$ and $v = v_0$ we are able to find values of the function $\Psi$ at each of the points of the grid.

\begin{figure}[htb]
\resizebox{.5\textwidth}{!}{\includegraphics{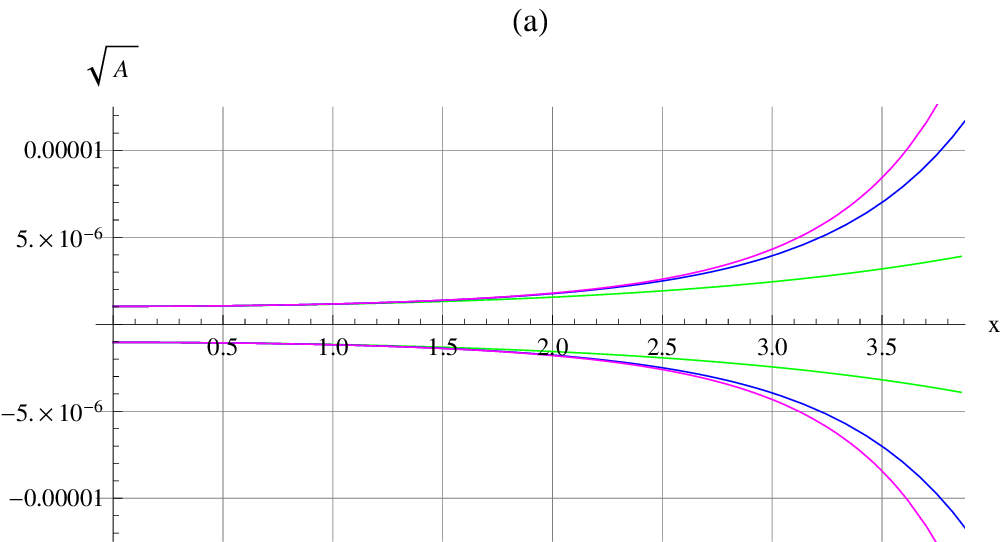}}
\resizebox{.5\textwidth}{!}{\includegraphics{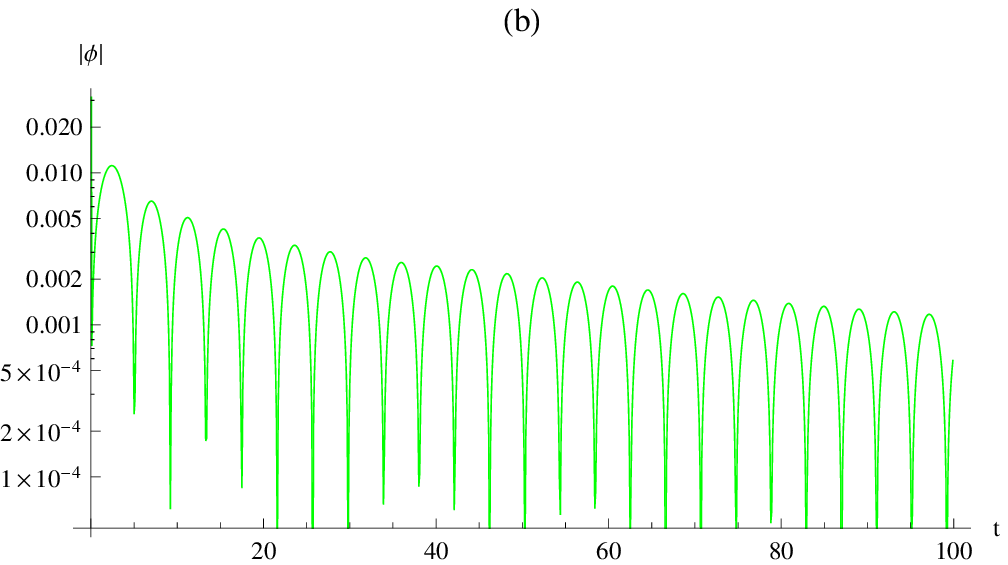}}
\\
\resizebox{.5\textwidth}{!}{\includegraphics{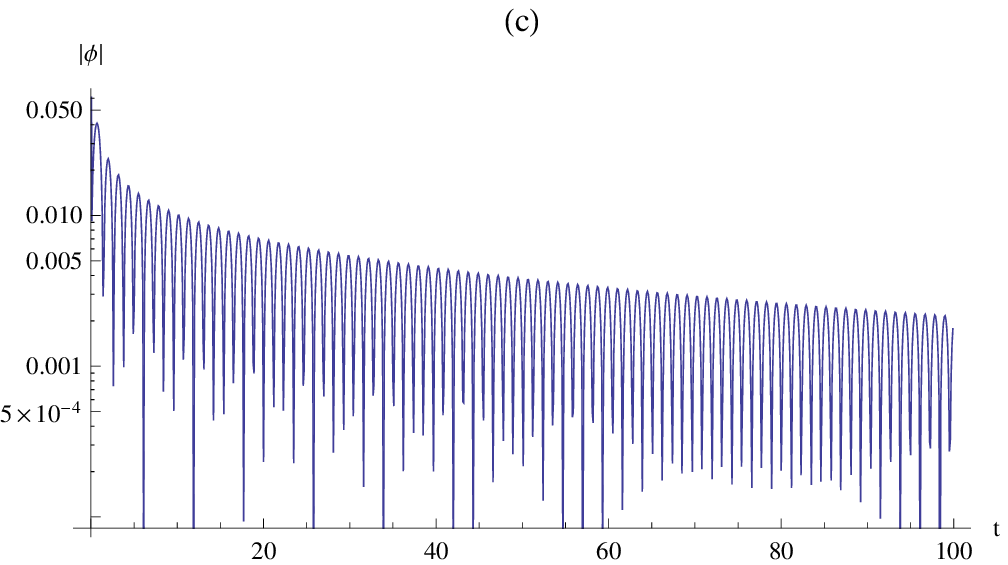}}
\resizebox{.5\textwidth}{!}{\includegraphics{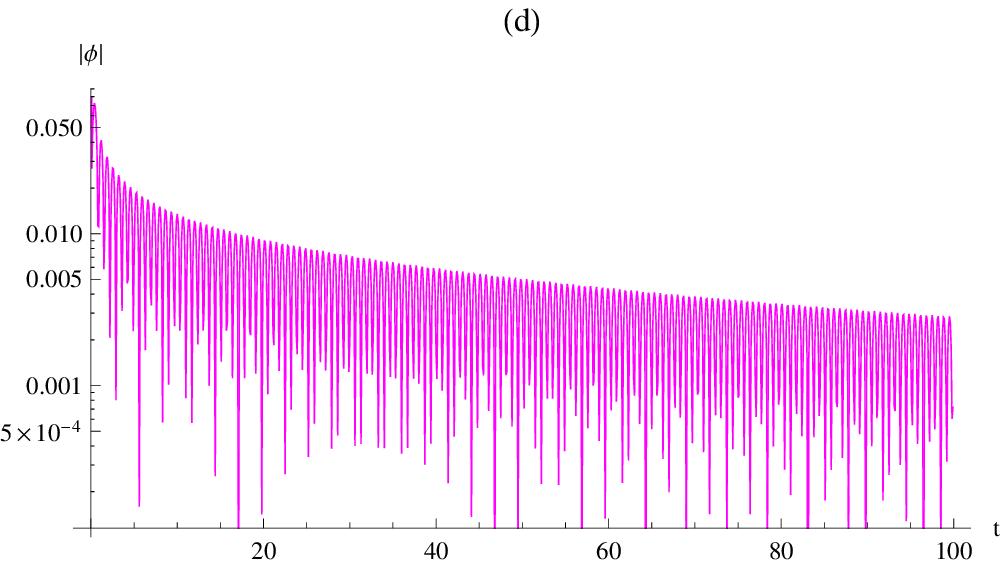}}
\caption{The form of the Laval nozzle for the spherically symmetric ($\ell=0$) massive scalar field (a): $m=0.8$ (green, 	
narrow), $m=2.84$ (blue), and $m=4.86$ (magenta, wide), and the corresponding time-domain profiles: (b) $m=0.8$, (c) $m=2.86$, (d) $m=4.86$.}\label{Fig_l0}
\end{figure}

\begin{figure}[htb]
\resizebox{.5\textwidth}{!}{\includegraphics{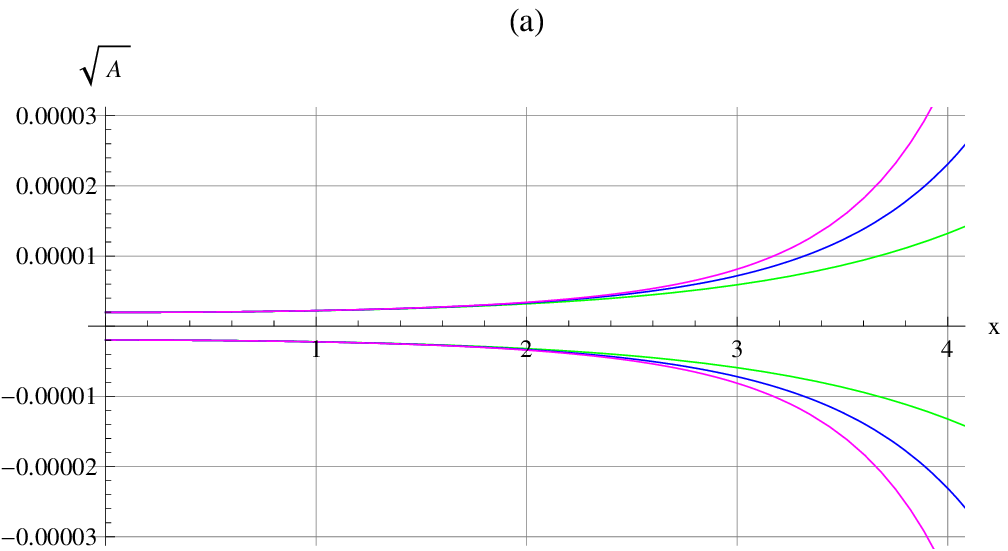}}
\resizebox{.5\textwidth}{!}{\includegraphics{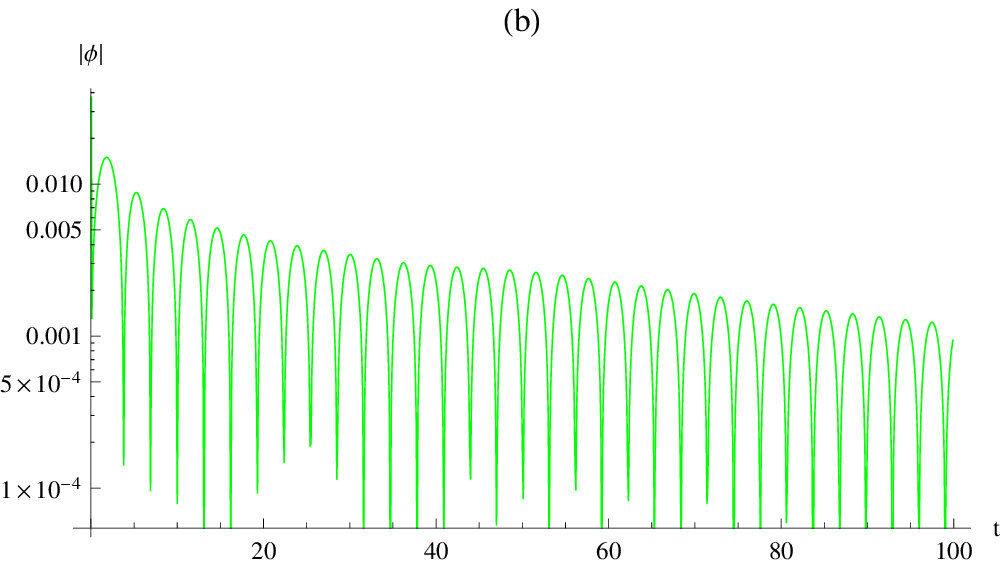}}
\\
\resizebox{.5\textwidth}{!}{\includegraphics{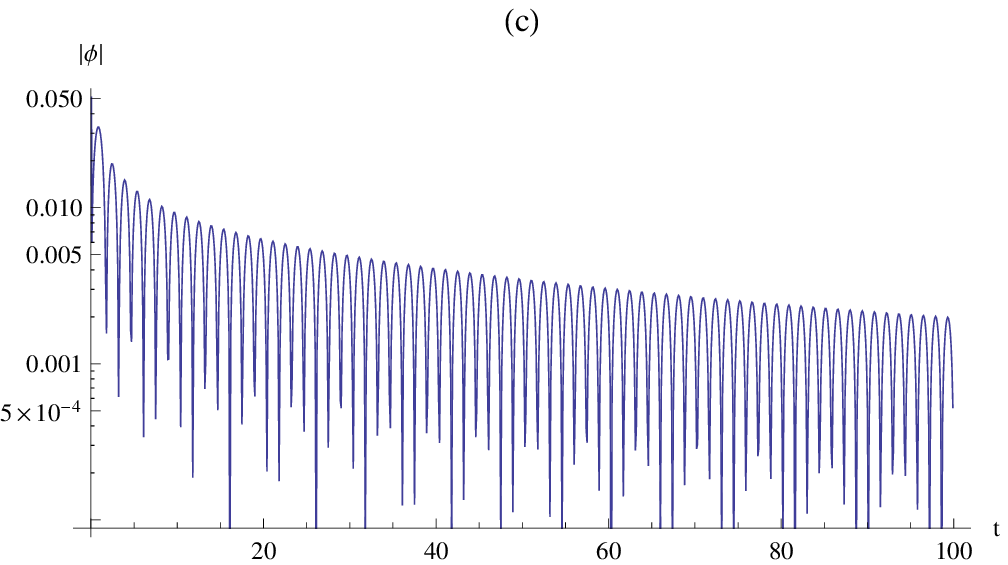}}
\resizebox{.5\textwidth}{!}{\includegraphics{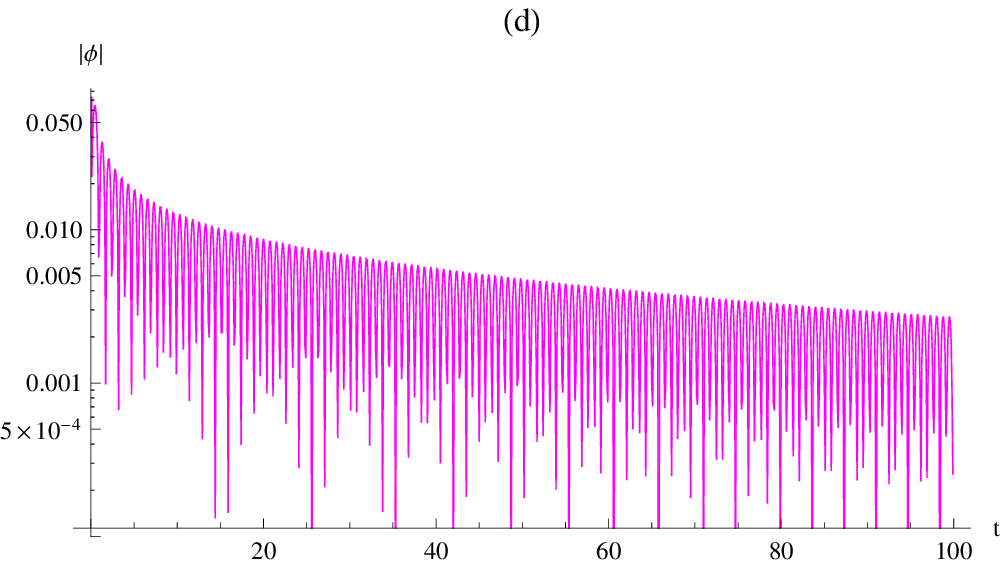}}
\caption{The form of the Laval nozzle for the $\ell=1$ massive scalar field (a): $m=1.06$ (green, 	
narrow), $m=2.30$ (blue), and $m=4.40$ (magenta, wide), and the corresponding time-domain profiles: (b) $m=1.06$, (c) $m=2.30$, (d) $m=4.40$.}\label{Fig_l1}
\end{figure}

On the figures (\ref{Fig_l0}) and (\ref{Fig_l1}) we show the forms of the nozzle that are acoustic analogs of the massive field with particular values of the mass for which nearly infinitely long-living oscillations exist. These oscillations called quasiresonances one can observe on the corresponding time-domain profiles. We see that in the tube of a particular form the decay rate of sound waves is almost zero for some tone which is an analogue of the quasiresonance of the massive scalar field.

Although we present here only the nozzles where the quasiresonances can be observed, the method described above can be used to construct an analogue for any finite mass of the in such a way that the sound waves in the nozzle will have the same behaviour as the massive scalar field in the background of the Schwarzschild black hole. From the figures (\ref{Fig_l0}) and (\ref{Fig_l1}) one can observe that the higher mass is, the quicker the nozzle cross-section grows, diverging at the end. However, as it was pointed out in \cite{23}, this does not lead to a problem with the presented model because of the freedom of the choice of the units of length. One can rescale the nozzle along the transversal axis in order to make the cross-section change as slowly as one wants. This change of the scale changes proportionally the frequencies of the sound in the nozzle.

\noindent

\section{Conclusion}
We have considered the Laval nozzle as an acoustic analogue of the massive scalar field in the background of the Schwarzschild black hole.
We presented the general method to determine the form of the Laval nozzle, such that the sound waves in it are described by a given effective potential, what can be used to study an analogue of black hole perturbations in a laboratory. The method can be used to obtain the forms of the nozzles, which are acoustic analogs of other spherically symmetric black holes. The acoustic analogs for perturbations of Reissner-Nordstr\"om(-de Sitter) black holes and their higher-dimensional generalizations, black strings, and Gauss-Bonnet black holes are of special interest. For some set of the parameters the higher-dimensional black holes and black strings suffer instability \cite{Gregory:1993vy,Gleiser:2005ra,Konoplya:2008au}, which in the corresponding nozzle can manifest itself as increasing of the sound amplitude.
It is clear that for a large amplitude of the sound waves cannot be described within the linear approximation and the considered analogue between the linear perturbations cannot be applied. Nevertheless, we believe that the consideration of different physical systems, which have linear instability in the same parametric region, could help us to understand better its nature.

\section*{Acknowledgments}
I would like to thank to my supervisor A.~Zhidenko for his help in preparing this paper for publication. This work was supported by Coordena\c{c}\~ao de Aperfei\c{c}oamento de Pessoal de N\'{\i}vel Superior (CAPES), Brazil.




\end{document}